\documentclass[linenumbers,trackchanges]{aastex701}
\usepackage{amsmath}
\usepackage{graphicx,subfigure}
\usepackage{gensymb}
\usepackage{appendix}
\usepackage{mathrsfs}

\usepackage[table]{xcolor}
\renewcommand{\textbf}[1]{#1}

\begin{document}

\title{On the importance of geometry in exoplanet irradiation : Implications for the day-night contrast}

\author[orcid=0000-0002-8095-2014]{Mradumay Sadh}
\affiliation{School of Mathematical and Physical Sciences and Astrophysics and Space Technologies Research Centre, Macquarie
University, 2109, NSW, Australia}
\email[show]{mradumay.sadh@students.mq.edu.au}  

\author[orcid=0000-0002-6603-9253]{Lorenzo Gavassino} 
\email{lorenzo.gavassino@gmail.com}
\affiliation{Department of Applied Mathematics and Theoretical Physics, University of Cambridge, Cambridge CB3 0WA, United Kingdom}

\begin{abstract}

The irradiance received by a spherical body or a planet close to a spherically symmetric source does not follow the point-sized source approximation and the inverse-square \textbf{variation of} irradiation \textbf{if} spherical symmetry is broken. In the penumbral zones of the planet, spherical symmetry of the star reduces to an axial symmetry. Our work aims to put forward a fundamental explanation, using energy conservation, to determine the variation of irradiance in the penumbral zone on a close-in planet where the point-sized source approximation fails. Consequently, we propose a numerical model that accurately predicts the irradiance within the boundaries of the penumbral zone and the fully-illuminated zone. Our analysis also corrects a previous study on exoplanet irradiation that violates energy conservation. We find that night-side illumination \textbf{partially explains the observed night-side temperatures on the planets considered; this reduces reliance on heat transport models} to explain the night-side temperature for \textbf{the few exemplar rocky close-in planets, namely K2-141 b, 55 Cancri e, TOI-561 b, TOI-431 b, and Kepler-10 b, that are discussed in this work. We provide improved day-night contrast temperatures, considering an airless scenario, and highlight the need for revisiting the heat transport models associated with atmospheric modelling of planets where the night-side illumination is significant.}

\end{abstract}



\section{Introduction}

\label{sec1}
\textbf{The inverse-square (IS) law is a consequence of the point-sized source approximation under spherical symmetry, which fails to be generally applicable to close-in, extremely irradiated systems or systems with very small semi-major axes.}
The irradiation effects between close stellar sources have mainly been an esoteric topic confined to the light curve analysis of binary star systems \citep{kopal1954photometric, wood1973reflection, wilson1990accuracy, huang2000effect}. \citet{kopal1954photometric}, in his seminal paper, derived a general theory of irradiation \textbf{specifically for the extreme effects that occur in close-in binaries where the components are distorted from a spherical shape, and  irradiation is affected by} reflection effects among the binary star system. With multiple discoveries of close-in exoplanets, starting almost three decades ago \citep{mayor1995jupiter}, the theory developed by \citep{kopal1954photometric} has not been extensively applied to exoplanetary systems, with only a few studies, for instance \citet{budaj2011reflection, nguyen2020irradiation}, utilising the model. \textbf{\citet{nguyen2020irradiation} used the model for the planet K2-141 b and concluded that the extreme geometry of this system makes an Si$\rm O_2$ atmosphere model feasible and also an easier observational target with JWST. 
Our work independently reproduces the same irradiance patterns and aims to broadcast the idea of the aforementioned extreme geometry by also applying our model to similar extreme systems.}
In recent years, \citet{carter2019irradiance,carter2024irradiation} have highlighted the importance of hyper illumination on close-in planets, \textbf{by improving upon the work of \citet{kopal1954photometric}. Their analysis shows that more than half of a close-in planet's surface is illuminated by its host-star. They also show that one of the previously used models, \textit{starry} \citep{luger2022analytic}, shows incorrect irradiance at the sub-stellar point on such planets. Despite of their improvements to previous works, their numerical solution in the penumbral zone does not agree with energy conservation when compared with the IS law. The inverse-square law serves as the benchmark against which comparisons should be made since previous attempts at calculating irradiance on close-in planets \citep{carter2019irradiance,2020MNRAS.499.1627S,luger2022analytic, carter2024irradiation} have faced problems with energy conservation repeatedly. \citet{carter2024irradiation} included the foreshortening of area elements on the star that were neglected by the \textit{starry} model \citep{luger2022analytic}. Thus, improving the associated modelling of such planets. We extend the accuracy further by proving an accurate and simpler explanation of why the IS law fails. Our objective is to focus not only on the geometry but also the reason why it plays a role on the irradiation received by the planet. Consequently, we build a fully consistent model using the fundamental understanding of radiative transfer and electrodynamics. We aim not only to calculate highly accurate irradiance values which have not been calculated hitherto in research, but also allude to the possibility of night-side illumination reducing the degeneracy in heat transport and atmospheric models of such planets. The limitations of the IS law naturally manifest when incorporating the night-side because it neglects the fact that some of the stellar surface from a large star illuminates the night-side.  
}

The simplicity of a star-planet system \textbf{without extreme gravitational distortion and reflection effects} makes the problem less intricate \textbf{than previous works on binary stars} and more applicable to a growing field of exoplanetary science. Our \textbf{approach is integrative, with an} aim to collate information from previous studies \textbf{through} an independent \textbf{model} originating from the most fundamental level. The \textbf{conceptualizing of hyper-illumination on planets very close to their stars started early after the discovery of 51 Peg b \citep{mayor1995jupiter}. }\citet{seager1998extrasolar} emphasised the importance of accurately modelling stellar irradiation on close-in extrasolar giant planets (EGPs). However, their approach relied on the inverse-square approximation. 
The inverse-square approximation has ubiquitously been used in previous studies to understand the thermal phase curves of exoplanets \citep{Madhusudhan2012,Farr2018,Haggard2018,Berdyugina2019,Kawahara2020,Heng2021,Teinturier2022}. In the following sections, we explain, with the help of geometry and radiative transfer, how the inverse-square law fails for one of the most extensively studied subsets of exoplanets, namely close-in exoplanets, and explore the astrophysical implications of this violation.


\section{The Foundation of our Method}\label{Appendix}


We begin with the basics of electrodynamics to set the foundation of our work. Under the conditions of vacuum and time-invariant radiation energy density \citep{jackson1999classical}, the Poynting theorem of energy conservation reduces to 
\begin{equation}\label{IOconservo}
    \rm \nabla \cdot S =0.
\end{equation}
Where S is the time-averaged Poynting vector. If we consider an arbitrary Gaussian 2D surface $\Sigma$, then equation \eqref{IOconservo} implies
\begin{equation}\label{Tucensrvi}
    \rm \oint_\Sigma S \cdot d {\Sigma} =0.
\end{equation}
We will use this formula to see where the inverse-square law is valid and where it is not, by appropriately choosing $\Sigma$.

\subsection{The method of the solid angle}\label{mathodangle}

Throughout the paper, we will focus, for simplicity, on a heterogenous binary system, comprising a star and a single planet. Both objects are approximated as perfectly spherical. In addition, since we are interested only in computing the energy that the star is transferring to the planet, we will treat the latter as a perfect absorber. In other words, we will not consider the contributions to \textit{S} due to emission and reflection of the planet (which are beyond the scope of the paper and would require us to model the atmosphere of the planet in detail).

\begin{figure}
    \centering
    \includegraphics[scale=0.35]{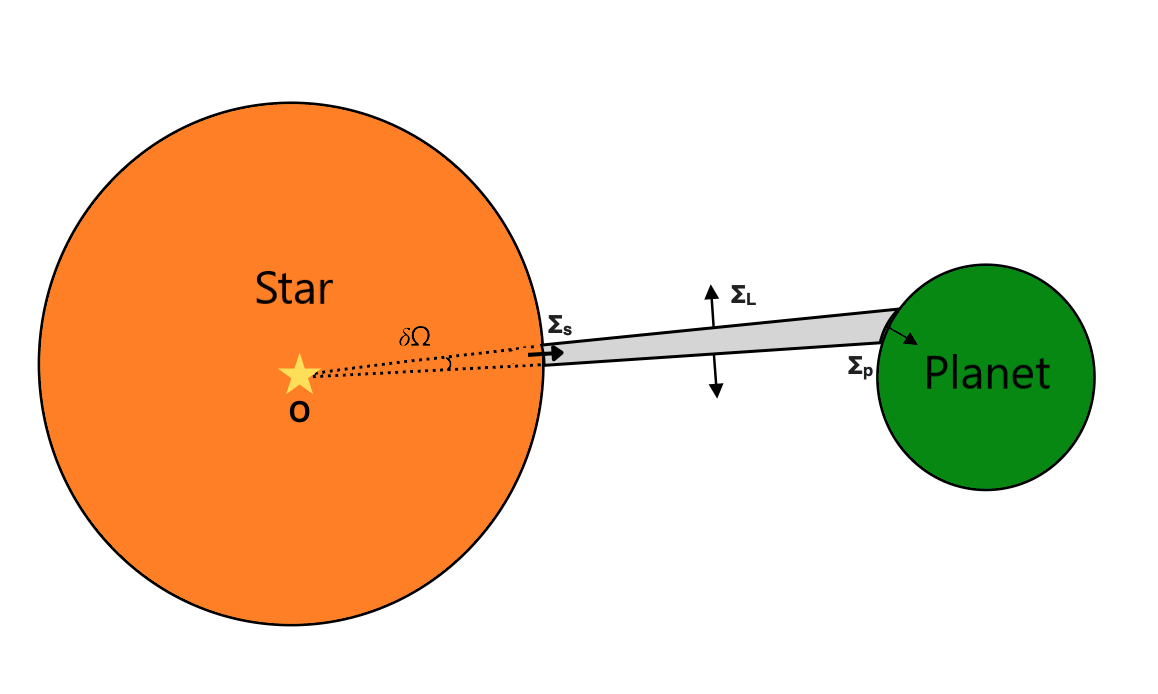}
    \caption{Schematic representation of the Gaussian 2D surface constructed in subsection \ref{mathodangle}. An infinitesimal solid angle $\delta \Omega$ is centred in the centre of the star and reaches the surface of the planet. Along its path, it ``cuts out'' a portion of space between the star and the planet whose boundary is the surface under consideration. The arrows in the picture are the normal unit vectors to the surface. The yellow star denotes the star as equivalent to a point-sized source in case of spherical symmetry.}
    \label{figurarrara}
\end{figure}

To probe the validity of the inverse-square law, we construct a Gaussian 2D surface as follows (see Fig. \ref{figurarrara}):
\begin{itemize}
    \item We consider an infinitesimal solid angle $\delta \Omega$ which is centered in the center of the star and points towards the planet, intersecting a small portion of its surface we call this intersection $\Sigma_P$ and we orient it from the star to the planet.
    \item  We call $\Sigma_S$ the portion of the surface of the star that intersects $\delta \Omega$ and we orient it from the star to the planet.
    \item We call $\Sigma_L$ the lateral surface of the solid angle crossing the vacuum between the star and the planet and we orient it outwardly. 
    \item $\Sigma_P \cup \Sigma_L \cup \Sigma_S$ is a connected Gaussian outwardly oriented surface which encloses a volume containing only radiation. This volume is the vacuum region between the star and the planet that intersects the solid angle. 
\end{itemize}
Applying the energy conservation condition \eqref{Tucensrvi} to $\Sigma_P \cup \Sigma_L \cup \Sigma_S$, we obtain
\begin{equation}\label{primuz}
      \rm \int_{\Sigma_P} S \cdot d {{\Sigma} } +     \int_{\Sigma_L} S \cdot d {{\Sigma}} +     \int_{\Sigma_S} S \cdot d {\Sigma}  =0.
\end{equation}

\noindent On the other hand, if $\rm L_b$ and $\rm R_s$ are respectively the bolometric luminosity and the radius of the star, we have (recalling the orientation conventions we imposed considering the outward nature of the flux from the star)
\begin{equation}\label{secunduz}
     \rm \int_{\Sigma_S} S \cdot d {\Sigma} = \dfrac{\delta \Omega}{4\pi}  L_b
\end{equation}
Therefore, equation \eqref{primuz} can be equivalently rewritten in the form
\begin{equation}\label{miraggioso}
    \rm \int_{\Sigma_P} S \cdot d {{\Sigma} } = - \dfrac{\delta \Omega}{4\pi}  L_b -  \int_{\Sigma_L} S \cdot d {{\Sigma}} \, .
\end{equation}
 We are going to show that, depending on the position of $\Sigma_P$, equation \eqref{miraggioso} might reduce to inverse-square law or might be in contradiction with it. 
\subsection{Case 1: Zero lateral contribution }
To find those regions, on the surface of the planet, where \eqref{miraggioso} reduces to the inverse-square law, we assume zero net flux contribution through ${\Sigma_L}$ ; this will occur when there is spherical symmetry in the source shape. \textbf{Therefore, the irradiance from the star will behave as if it is coming from a point-sized source centred at the stellar centre (denoted by O in Fig. \ref{figurarrara}). The approximation of a spherical star is valid for all stars whose shape is not distorted by a massive secondary stellar component.}
 Mathematically,
\begin{equation}\label{percheradio}
       \rm \int_{\Sigma_L} S \cdot d {{\Sigma}} =0,
\end{equation}
If we insert \eqref{percheradio} into \eqref{miraggioso} and use the definition of solid angle,
\begin{equation}
    \rm \hat{r} \cdot {\Sigma}_P = d_c^2 \, \delta \Omega,
\end{equation}
where $\rm \hat{r}$ is the unit radial vector \textbf{of the spherical coordinate system, originating from the centre of the star} and $\rm d_c$ is the distance of $\Sigma_P$ from the centre of the star. We obtain the inverse-square law for the energy current:
\begin{equation}\label{maegrande}
    \rm S_P = \dfrac{L_b}{4\pi d_c^2}.
\end{equation}
Finally, if we call $\rm \hat{\bm{\nu}}$ the unit normal vector to $\Sigma_P$ and use it to define the geometric quantity
\begin{equation}
    \rm \cos i := \hat{\bm{r}} \cdot \hat{\bm{\nu}},
\end{equation}
then, using \eqref{miraggioso} and \eqref{maegrande}, we obtain the inverse-square law for irradiance:
\begin{equation}\label{Iinverseqsqasqd}
    \rm I := \lim_{\delta \Omega \rightarrow 0} \bigg[ \dfrac{1}{\Sigma_P}  \int_{\Sigma_P} S \cdot d {{\Sigma} } \, \bigg] = \dfrac{L_b \cos i}{4\pi d_c^2}.  
\end{equation}
We note here that the angle \textit{i} is different from the latitude ($\lambda$) or the zenith angle (which assumes the star on the celestial equator) of the point-sized source (O) as seen from the planet, which is used in other studies, for instance \citet{guillot2010radiative,maurin2012thermal}, assuming plane-parallel approximation. In case of the plane-parallel approximation, \textbf{with incident stellar rays being parallel to each other,} \textit{i} reduces to $\lambda$ (refer to appendix \ref{general}). However, we ignore the plane-parallel approximation in our analysis and use the general equation above.

\subsection{Case 2: Beyond the critical point of symmetry} \label{violation}

The critical latitude of symmetry comes when we transition from the fully-illuminated latitudes on the planet to the zone of the partially eclipsed latitudes or the penumbral zone. \textbf{It is in this region that an observer on the planet does not see a complete circular disk of the star and instead observes a disk sector, breaking the polar symmetry on the spherical star. In this work, we refer to the polar angle as measured along the perpendicular direction to the orbital plane of the system.}
Equation \ref{miraggioso} will be in contradiction with the inverse-square law in the penumbral zone. We explain the reason for that in this section. The region within the interior tangent planes of the stellar and planetary spheres will always obey the inverse-square variation since the star follows spherical symmetry in this region. As the star is eclipsed, the spherical symmetry becomes solely axial. The solutions are therefore approximated through Legendre polynomials \citep{kopal1954photometric, carter2024irradiation}. The geometry and mathematics associated with this explanation is discussed in detail in the work of \citep{kopal1954photometric}, and therefore, not repeated here except a few important equations in the appendices \ref{general} and \ref{limb}. The major difference in our equations lies in the consideration of the primary and secondary sources of the binary system. They have developed their equations considering the irradiation from the secondary or smaller component onto the primary. In contrast, we have only one radiating source, the star, which is the primary source irradiating the smaller planetary or secondary component.
\citep{kopal1954photometric} noted that there is no closed-form solution to the irradiance in the penumbral zone, therefore we have used numerical integration of the elliptic integrals in our calculation.

We refer to the intersection point of the common interior tangents, seen as the black lines in Fig. \ref{occhio}, with the planetary circle (sub-stellar longitude of the planet) as the critical point of symmetry ($\rm \lambda_{CPS}$). The inverse square law will be violated beyond the critical point of symmetry but within the region marked by the exterior common tangents, which are the grey lines seen in Fig. \ref{occhio}.
The exterior common tangents mark how far the day-night terminator will extend beyond the poles, towards the night-side of the planet. The dark-green belt in Fig. \ref{occhio} marks the penumbral zone where we expect to see deviations from the spherical symmetry of the star. Beyond the terminator limit (T), or in the umbral zone, the flux is naturally zero.

\subsubsection{An intuitive explanation}

\begin{figure}[h!]
    \centering
    \subfigure[]{
    \label{occhio}
        \includegraphics[scale=0.15]{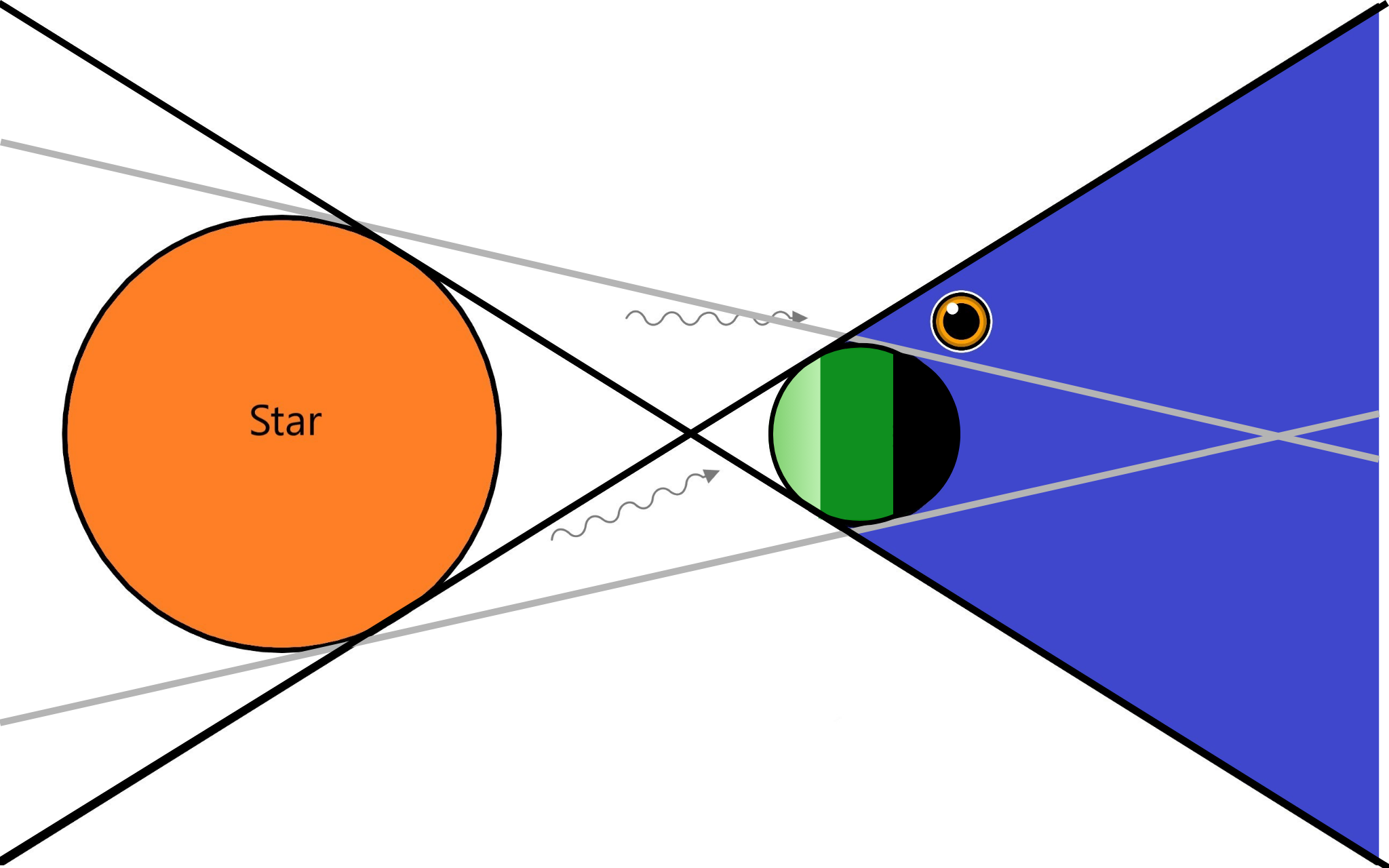}
        
    }
    \hfill
    \subfigure[]{
    \label{tangent}
        \includegraphics[scale=0.15]{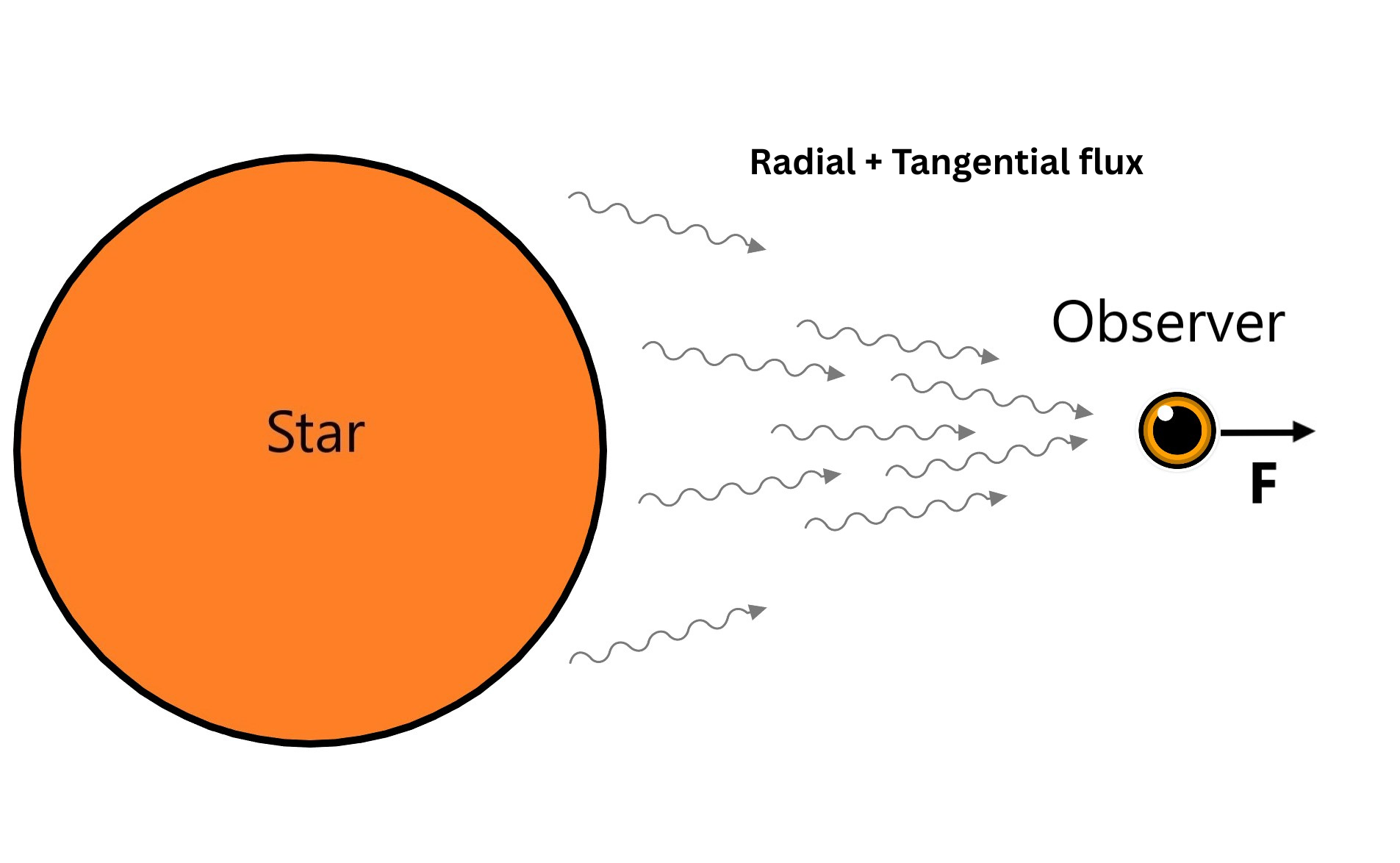}
        
    }
    \caption{Fig. (a) shows geometry of the common tangents to the stellar and planetary circles. The blue region is the set of all those positions from which the star is not entirely visible. An observer (represented in our case by the eye) sitting on an arbitrary blue point would see a partially eclipsed star and hence a non-zero tangential component of the flux. This opens the door to possible violations of the inverse-square law on the dark green region of the planet. However, one should note that the violations are only expected where the flux from the star is non-zero. For instance, in the black region on the planet beyond the exterior common tangents, the flux is naturally zero. Fig. (b) shows the observer seeing the complete star, which receives zero net tangential component of the flux due to polar symmetry, leaving \textit{S} purely radial.}
    
\end{figure}

The time-averaged Poynting vector (\rm \textit{S}) exhibits two components, namely radial and tangential (Fig. \ref{tangent}). In the case of axial symmetry, multipole expansion of Legendre polynomials governs the radial and tangential components \citep{jackson1999classical}. Here, we simply write the components as,

\begin{equation}\label{tangential}
    \rm S= \alpha S_{r} \hat{r} + \beta S_{\theta} \hat{\theta} ,
\end{equation}

Where $\alpha$ and $\beta$ are coefficients of each component, which depend on geometrical parameters and boundary conditions. The flux emanating through $\Sigma_S$ is always radial. The net flux entering and exiting $\Sigma_L$ will never be radial \textbf{due to its area vector being perpendicular to the radial vector $\hat{r}$}, and will be zero when the tangential flux entering will exit symmetrically. The flux received by $\Sigma_P$ can be both radial or tangential depending on the latitude. Due to polar asymmetry \textbf{in penumbral zones}, the non-zero flux through the lateral surface will exit through $\Sigma_P$; this causes the additional irradiance that IS law misses. 

The IS law always measures radial flux. The radial component is dominant for lower and moderate latitudes and it decays with cosine variation of the angle \textit{i}, as shown in equation \ref{Iinverseqsqasqd}, and goes to zero near poles. When we consider the star with its finite size, the tangential component comes into picture beyond the critical point of symmetry ($\rm \lambda_{CPS}$) and between the region governed by the interior and exterior tangent planes. The net tangential component rises beyond $\rm \lambda_{CPS}$ due to polar asymmetry of the star, slowing the cosine flux decrement with latitude. It reaches its maximum value at the poles due to maximum asymmetry about the orbital axis ($\theta=0$). Beyond the poles, the only remaining (tangential) component now reduces as the eclipsed area reduces till the terminator limit (T).

\section{\textit{InstellCa} -- A numerical code}

\begin{figure}[h!]
    \centering
    \includegraphics[scale=0.35]{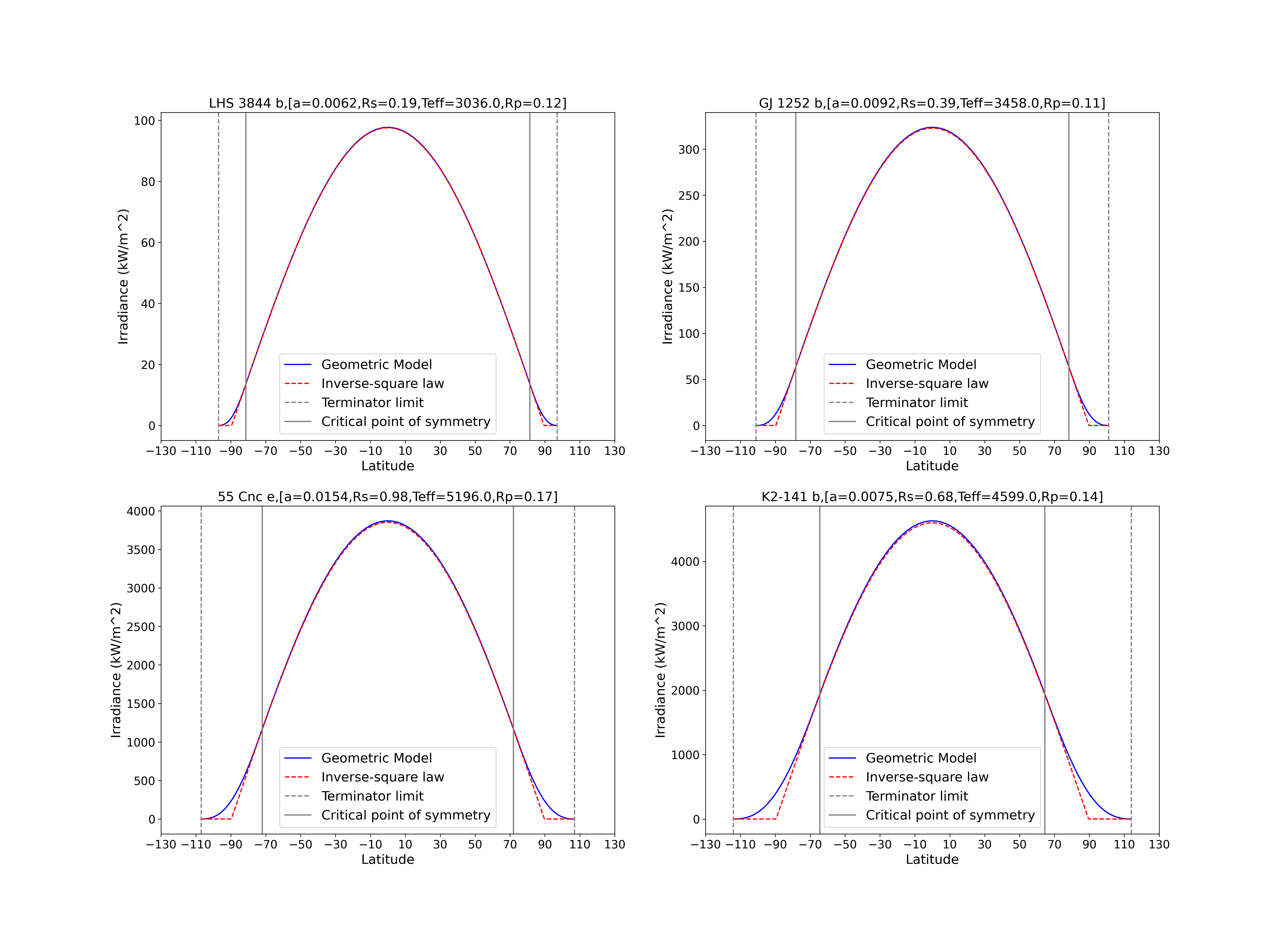}
    \caption{An \textit{InstellCa} plot showing irradiance received across the sub-stellar longitude of the exoplanets LHS 3844 b, 55 Cancri e, GJ 1252 b, and K2-141 b. The title shows the parameters used by the code (Table \ref{Exo-parameters}). We see largest manifestation of this effect for K2-141 b due to the larger visible angular size of the star ($\sim$ 25 degrees). For all planets, the numerical model overlaps with the inverse-square law for the fully illuminated zone. The slight difference between both approaches in the fully-illuminated zone is attributed to numerical error of less than one percent. In the penumbral zone, deviations are seen as the critical point of symmetry is crossed. The irradiance naturally falls to zero beyond the day-night terminator limit. }
    \label{example}
\end{figure}

We have developed a Python-based numerical model\footnote{The code \textit{InstellCa} has been made openly available at \href{Code}{https://github.com/Mradumay137/InstellCa}.
} for calculating irradiance on exoplanets, including the penumbral zones. The model is created considering the star as a spherical body and provides the irradiance pattern at all irradiated latitudes of the sub-stellar longitude on the \textbf{spherical} planet. Fig. \ref{example} shows the irradiance pattern for the well-studied close-in planet 55 Cancri e \citep{mcarthur2004detection, demory2011detection} and rocky, airless exoplanets K2-141 b \citep{malavolta2018ultra,zieba2022k2}, GJ 1252 b \citep{shporer2020gj,crossfield2022gj}, and LHS 3844 b \citep{vanderspek2019tess,kreidberg2019absence}, using parameters from the Encyclopaedia of exoplanetary systems \citep{schneider2011defining}, which are listed in table \ref{Exo-parameters}. We note a significant difference from the IS law near the poles of the planet, as predicted by theory in the previous section. The day-night terminator, instead of being at the poles, extends further towards the night-side. The larger the angular size of the star (see Table \ref{Exo-parameters}), the further the terminator extends, increasing the extent of the deviation.
\textbf{The irradiance difference from the IS law, in the penumbral zones, rises with the angular-size of the host-star as seen from the planet. Therefore, an angular size of 5 degrees will extend the terminator limit by approximately 5 degrees beyond the poles. }
Our numerical irradiance calculations have an excellent accuracy of 99.5 percent when compared with the IS law; this corresponds to an error of $<1$ ppm of $\rm I_{star}$. \textbf{This error, which is entirely numerical in origin, can be reduced further if the latitude arrays are made more densely spaced and close to a continuous variable. However, this increases the computation time exponentially.} A previous study on K2-141 b \citep{nguyen2020irradiation} predicted its irradiance pattern using the \citet{kopal1954photometric} model, which \textbf{differs by less than one percent from our calculated values} in Fig. \ref{example}. This serves as an independent verification of our approach with the standard method developed by \citet{kopal1954photometric}.

We also note that our code includes provisions for incorporating the effects of linear limb darkening and we have used linear limb darkening (refer to Appendix \ref{limb}) in Fig. \ref{example}. However, we find that limb darkening does not appreciably change the irradiance in both penumbral and fully-illuminated zones.

\begin{table}[]
    \centering
    \begin{tabular}{|c|c|c|c|}
    \hline
         Exoplanet & Semi-major axis (AU) & Host-star radius ($ \rm R_{\odot})$ & Planetary radius (Earth radii) \\
         \hline

         LHS 3844 b & 0.0062 & 0.19 & 1.30  \\
         
         GJ 1252 b & 0.0092 & 0.39 & 1.19  \\

         55 Cancri e & 0.0154 & 0.98 & 1.88  \\

         K2-141 b & 0.0075 & 0.68 & 1.50  \\

    \hline
    \end{tabular}
    \caption{The parameters obtained from the Encyclopaedia of exoplanetary systems \citep{schneider2011defining},\textbf{ which in turn obtained it from the following sources \citep{vanderspek2019tess, shporer2020gj,bourrier201855,crida2018mass, bonomo2023cold}} for all planets shown in Fig. \ref{example}. }
    \label{Exo-parameters}
\end{table}

\subsection{Accordance of results with theory}

\begin{figure}[h!]
\centering

\includegraphics[scale=0.5]{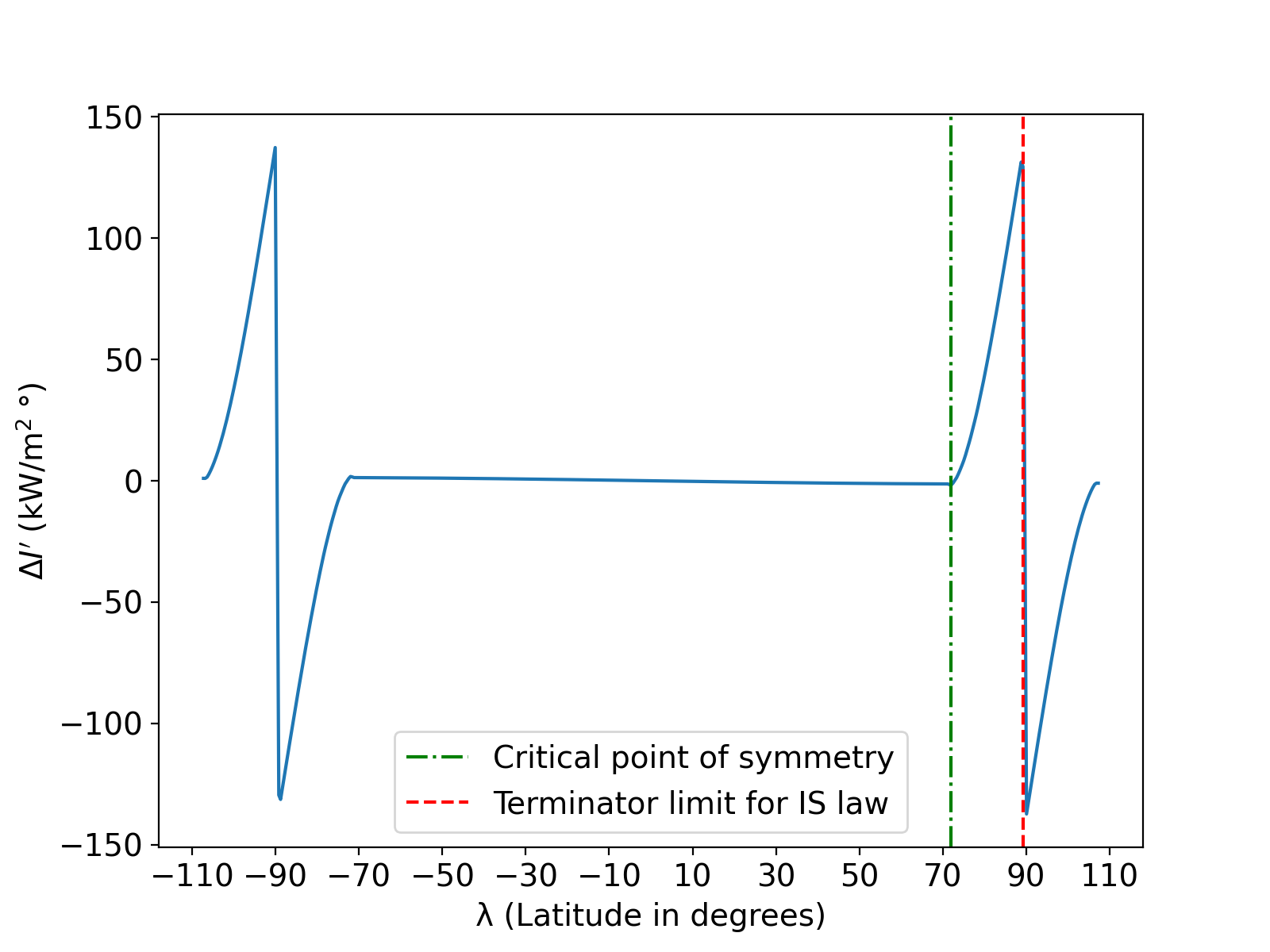}
\caption{Figure shows the derivative of the deviation from the IS law for the planet 55 Cancri e. The deviation from the IS law rapidly increases after the critical point of symmetry is crossed, implying consistency of the numerical results with the theoretical prediction. The red dotted line indicates the day-night terminator according to the points-sized source approximation, which is close to the poles. 
}
\label{proof}
\end{figure}

To confirm whether our derivation from basic radiative transfer is consistent with our numerical model, we compute the derivative of the difference in the two approaches with respect to the latitude ($\lambda$) on the planet. Mathematically,
\begin{equation}
    \rm \Delta I(\lambda)' = \frac{\mathrm{d}}{\mathrm{d}\lambda}\!\left( I_{NM} - I_{IS} \right)
\end{equation}
Where NM stands for the numerical model and IS implies the inverse-square law. We plot this in Fig. \ref{proof} to examine the variations of this difference. Since the critical point of symmetry ($\lambda_{\rm CPS}$) decides where the inverse-square law fails, we create a vertical (green) reference line for comparison.
We see excellent agreement in Fig. \ref{proof} for the prediction of the critical point of symmetry between two independent approaches; the theory described in section \ref{violation}, which suggests the critical point based on energy conservation, and our code, which calculates it using numerical integration.

\citet{carter2024irradiation} make a small but significant mistake in estimating the intersection angle of the observer's horizon with the host-star apparent disk. 
They assume the following limit for their irradiance integral (\textit{equation 19} of \citet{carter2024irradiation}),

 \begin{equation}\label{wrong}
    \rm \theta_{lim} = \pm \Bigg[\alpha - \cos^{-1}{(\frac{\rho \cos\alpha}{R_s})}\Bigg]
\end{equation}



Using equation \ref{wrong} underestimates the irradiance, which is consequently seen in \textit{Fig. 9} of \citet{carter2024irradiation}, where we see that the difference (the cyan curve in \textit{Fig. 9}) of their numerical model with the plane-parallel approximation becomes negative in PZ1; this is in clear violation of energy conservation. We provide the correct equation in Appendix \ref{limb}.

Overall, the maximum difference this creates in PZ1 is nearly 20 ppm of $\rm I_{star}$. An intensity difference of this magnitude is large and often equivalent to the secondary eclipse depth (23 ppm) in thermal phase curves of highly irradiated planets \citep{malavolta2018ultra}.
Moreover, a difference of 20 ppm at the poles, which is a crucial region for possible climate states \citep{kilic2017multiple} on a planet, might therefore be imperative to correct.

\section{Astrophysical Implications}

Dissimilar to typical phase curve studies on this topic \citep{Madhusudhan2012,Farr2018,Haggard2018,Berdyugina2019,Kawahara2020,Heng2021,Teinturier2022}, we choose to focus on a highly-simplified airless planet with no reflection off the surface; this is done to emphasise the importance of this geometric effect without the inclusion of atmospheric heat transport phenomena that lie beyond the scope of our current work. The geometric effects, in most cases, cause a faint night-side flux that is difficult to be detectable even with high precision measurements \citep{zieba2022k2}.
Moreover, we note that the anomalous phase offsets noted in multiple studies \citep{hammond2017linking,kreidberg2019absence,mansfield2023revealing} in thermal phase curves of highly irradiated planets cannot be explained under the phase-symmetric conditions we have assumed. However, incorporating our numerical model into a comprehensive atmospheric analysis may yield more informative outcomes. 

\subsection{Day-night contrast and the importance of night-side illumination}

In this section, we explore the astrophysical implication of our assumed model for a close-in airless planet with zero albedo. 55 Cancri e is an interesting example, since it is well-studied and puzzling \citep{demory2011detection, demory2016map, hammond2017linking, hu2024secondary,patel2024jwst}. \citet{hammond2017linking} noted that the large day-night contrast \textbf{(1300 $\pm$ 670 K) and large phase offset $(41 \pm 12)^\circ$measured by \citet{demory2016map}   in the \textit{Spitzer IRAC}} thermal phase curves of 55 Cancri e posed a puzzle. Essentially, they were posing contradictory interpretations about heat distribution and transport on the planet, casting doubts on the associated presence or absence of atmosphere. \citet{mercier2022revisiting} showed that the phase offset is smaller than the value previously modelled, reinforcing the idea of it being a lava planet with a tenuous or absent atmosphere \citep{meier2023interior}. Thus, the day-night contrast is a crucial piece of the puzzle.

Here we calculate, for 55 Cancri e, the fractional day-night contrast $A_T$ used in phase curve amplitude estimation of exoplanets \citep{komacek2016atmospheric,koll2022scaling,mansfield2023revealing}. 

\begin{equation}
    A_T = \frac{ T_{day} - T_{night}}{T_{day}}
\end{equation}

\noindent For the case of an airless planet, the hemisphere-averaged equilibrium temperature, for the hemisphere facing the star, can be calculated through the irradiation received by it:

\begin{equation}
    \rm I_{dayside} = \frac{\int^{\pi/2}_{0} I d\lambda}{\pi/2} = 2437 kW/m^2
\end{equation}

\noindent Therefore,

\begin{equation} \label{eqt}
    \rm T_{day} = \Bigg[ \frac{I_{dayside}}{ \sigma} \Bigg]^{1/4} = 2560 \degree K
\end{equation}

\noindent We note that this is nearly 23 K higher than what was estimated by \citet{patel2024jwst} due to the underestimation of irradiance, by the IS law used by them, near the poles. \textbf{This small difference stands far below the JWST uncertainty of 88 K for the planet \citep{hu2024secondary}.} Therefore, our model has minimal practical implications for the hemisphere facing the star. We also note here that this average automatically accounts for the heat redistribution factor \textit{f} = 2/3 \citep{koll2022scaling}, corresponding to an airless planet.

In a similar manner, calculating the hemisphere-averaged equilibrium temperature for the hemisphere facing away from the star will yield:

\begin{equation}
    \rm T_{night} = 691 \degree K
\end{equation}
\textbf{
\noindent Where we have assumed the integrand as,}

\[
\rm I =
\begin{cases}
\rm I, & \text{if } \pi/2 < \lambda < T,\\
0,  & \text{if } T<\lambda < \pi.
\end{cases}
\]

\noindent Therefore,

\begin{equation} \label{bench}
    \rm A_T^{airless} = 0.73
\end{equation}

Contrary to the day-side, our night-side irradiation temperature has significant implications, since the studies following the IS law ignore this. 

The observed brightness temperatures for 55 Cancri e, on the day and night side, will be a result of different values of $T_{day}$ and $T_{night}$ according to how heat is conducted through geological factors \citep{meier2023interior} and convected or re-radiated through atmospheric factors \citep{hammond2017linking}, which lead to lower values of \textit{f} from the maximum (2/3). However, equation \ref{bench} provides the benchmark for the theoretical upper limit of $A_T$, since heat redistribution will transport energy across the planet, reducing $A_T$ to $\sim$ 0.5, as evident in the works of \citet{demory2016map,mercier2022revisiting}, depending on the atmospheric model considered. Our model limits the amount of heat transported to the night-side since it suggests the presence of a baseline temperature due to night-side illumination, which is ignored in all exoplanet phase curve analyses. Consequently, the theoretical maximum fractional day-night temperature ($A^{airless}_{T}$) is assumed unity in thermal phase curve studies \citep{showman2002atmospheric,komacek2016atmospheric,mansfield2023revealing}, since $\rm T_{night}$ is assumed zero. 

Airless planets are not mere mathematical fiction. The planet GJ 1252 b (Fig. \ref{example}) has been reported to have no atmosphere \citep{crossfield2022gj} with its observed day-side temperature (1410 K) matching its equilibrium temperature (1399 K) within uncertainties. An extreme example is the planet K2-141 b (Fig. \ref{example}). We estimate a night-side temperature of 852 K for this planet using purely geometric effects of irradiation. The night-side temperature for K2-141 b is high enough to be detected without the presence of significant atmospheric effects and recorded to be $956_{-556}^{489}$ K \citep{zieba2022k2} , which can likely be explained solely using geometric effects without the need for incorporating heat transport phenomena, strengthening the case for K2-141 b to be a planet with little-to-no atmosphere \citep{zieba2022k2}.
 
Close-in or ultra-short period (USP) planets \citep{demory2016map, kreidberg2019absence, zieba2022k2,crossfield2022gj, teske2025thick} are highly pertinent for exoplanet climate studies and we intend to complement the atmospheric models, for the atmospherically anomalous USPs akin to 55 Cancri e, TOI-561 b \citep{teske2025thick}, Kepler-10 b \citep{bonomo2025depth} and TOI-431 b \citep{monaghan2025low}, with geometric effects to constrain the possible heat transport scenarios to the night-side and accurately model the presence, absence, or tenuity of an atmosphere. \ref{tab:placeholder}

\begin{table}[]
\centering
\begin{tabular}{|>{\columncolor{white}}c|>{\columncolor{green!20}}c|>{\columncolor{cyan}}c|}
\hline
Planet & Day-side $T_{d,max}$ (K) & Night-side $T_{n,max}$ (K) \\
\hline
K2-141 b & 2686 & 854 \\
\hline
55 Cnc e & 2560 & 691 \\
\hline
TOI-561 b & 2887 & 851 \\
\hline
TOI-431 b & 2415 & 650 \\
\hline
Kepler-10 b & 2825 & 763 \\
\hline
\end{tabular}
\caption{The table shows calculated hemispheric maximum temperatures, according to the definition $T_{d,max}$ by \citet{teske2025thick} for 5 extreme planets where we expect to find the largest extent of deviations in the night-side temperatures from the IS law. The temperatures are representative of a blackbody (planet) re-radiating all radiation it has received.
The third column with cyan background indicates our theoretically predicted maximum temperatures which are divergent from the IS law, since it predicts zero irradiance at the night-side, whereas the light green background in the second column is used to indicate that our day-side temperatures are in close agreement with the IS law based measurements. }
\label{tab:placeholder}
\end{table}


\subsection{Considering non-zero Bond albedos}

\textbf{In this section, we briefly explore the effects of assuming a non-zero albedo on the derived maximum day-side temperatures. We see that all USPs considered show observed day-side temperatures that can be explained, in principle, by a Bond albedo ($\alpha$) in the range $\alpha \in$ (0.65--0.8). However, atmospheric attenuation and other complex phenomena might influence this simplistic explanation of the planets considered. \citet{zieba2022k2} estimated a geometric albedo of 0.28, corresponding to a Bond albedo of 0.42 for K2-141 b, whereas \citet{hu2015semi} modelled the Bond albedo for Kepler-10 b to be higher than 0.8, giving weight to the hypothesis of such high albedos explaining the observed temperatures.
However, other studies, for 55 Cancri e and TOI-561 b, have modelled the day-side brightness temperatures assuming zero albedos \citep{hammond2017linking,teske2025thick}. }

\begin{figure}
    \centering
    \includegraphics[scale=0.5]{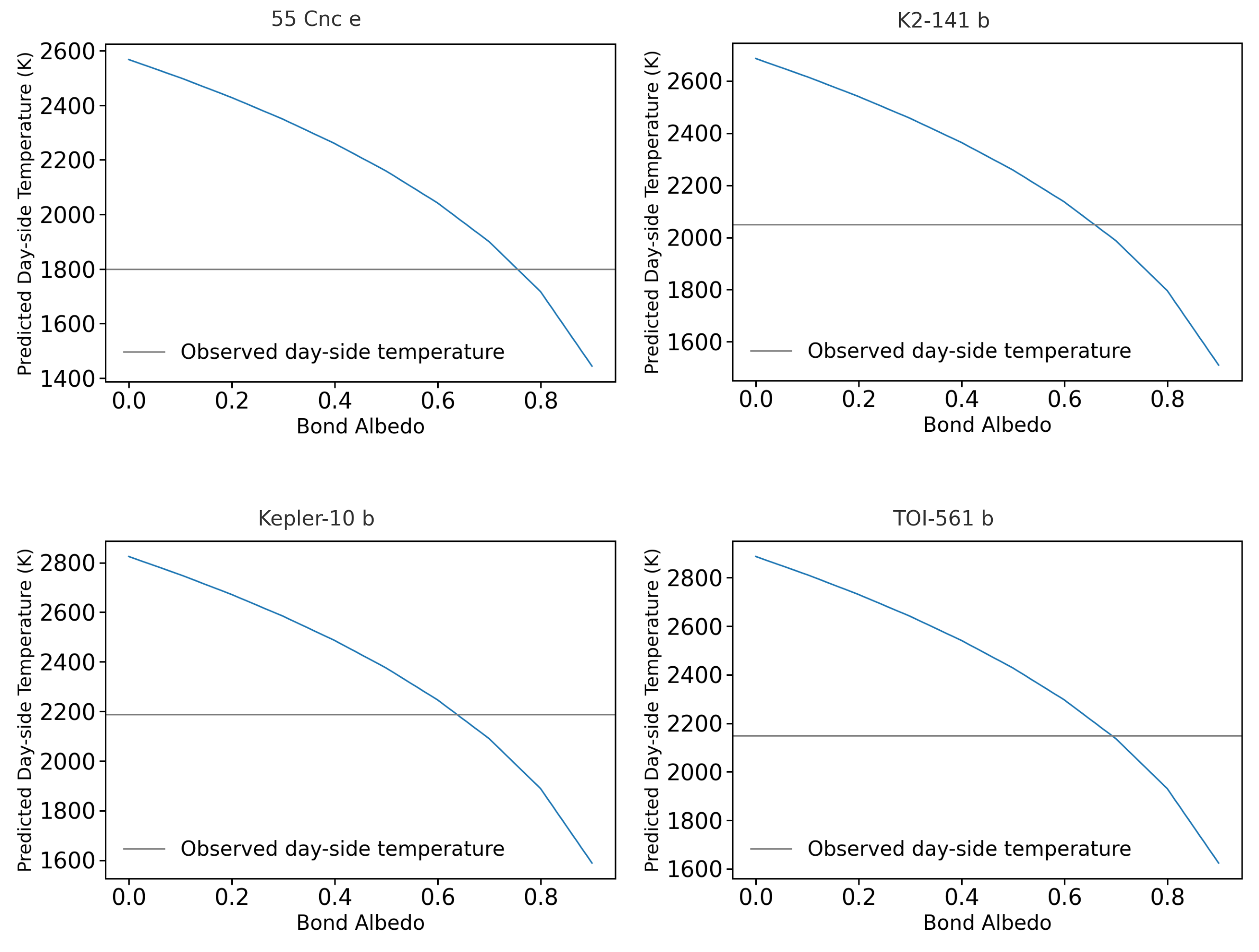}
    \caption{Variation of the predicted day-side temperature, which is concordant between our approach and the IS law, with multiple Bond albedo values. The observed brightness temperatures are taken from light curve analysis of each planet respectively \citep{hu2024secondary,zieba2022k2,bonomo2025depth,teske2025thick}.}
    \label{fig:placeholder}
\end{figure}

\section{Conclusion and Future Work}

We have attempted to explain the limits of the inverse-square law of radiation, using geometry and fundamentals of radiative transfer. Instead of depending on the emergent assumption \textbf{of the IS law always being valid}, we have purposefully chosen a first-principles approach; the concurrence of our numerical results with theory is a natural consequence of the simplicity of the geometry and physical laws involved. Moreover, we have also shown that stellar and planetary geometry may alter how night-side brightness temperatures are modelled for planets \textbf{that may or may not possess} atmospheres, which could help in the analysis of planets with debatable atmospheres. Although the night-side irradiation temperatures produce a very weak night-side flux without atmospheric effects, extreme cases like K2-141 b, with the terminator extending to 115 degrees from the sub-stellar point, help show how planets with tenuous atmospheres and minimal heat transport can also show borderline detectable night-side flux, in the range of 10--20 ppm, with JWST capabilities \citep{kreidberg2025first}.

\textit{InstellCa} can be used for all planets in the current exoplanet databases \citep{schneider2011defining, akeson2013nasa} with the airless planet assumption. Future work will involve adding reflection and atmospheric effects to generate model thermal phase curves of multiple well-studied close-in exoplanets for enabling comparison with observational data. The addition of atmospheric analysis will broaden the scope of our work to other close-in planets with either thin or thick atmosphere and allow benchmarking through future JWST observations \citep{kempton2024transiting}; this will also enable the measurement of day-night temperature contrast on lava-planets (For example, TOI-2431 b \citep{tacs2025earth}).
Such comprehensive models can assist the software \citep{knuth2017exonest} that wish to address these effects in future. Since our work fundamentally shifts the day-night terminator, it may have profound implications for terminator habitability \citep{lobo2023terminator} that we aim to explore in subsequent works.

\begin{acknowledgements}
We would like to express our sincere gratitude to Dr. Joanne Dawson, Prof. Mark Wardle, and the AAS reviewer who provided highly constructive feedback that amplified the elegance and relevance of the article for the astronomical community.
\end{acknowledgements}


\appendix

\section{General equation for the inverse-square law}\label{general}

\begin{figure}[h!]
    \centering
    \includegraphics[scale=0.3]{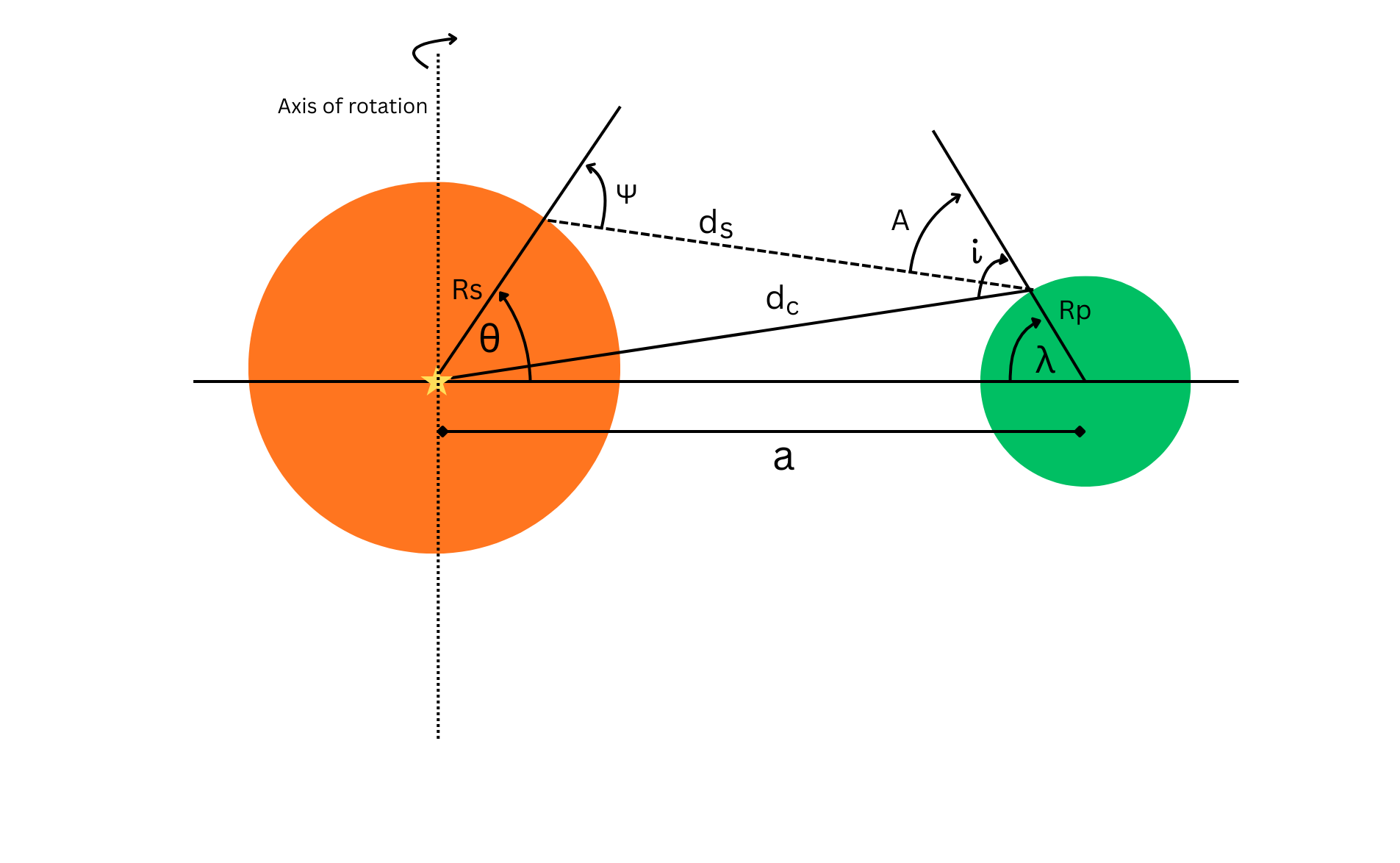}
    \caption{A diagram showing the angles and distances considered for a star-planet system with semi-major axis \textit{a}. The yellow star represents point-sized source approximation, corresponding to the inverse-square law. The axis of rotation is shown as the reference of azimuthal symmetry.} 
    \label{geo}
\end{figure}

To compare the numerical model with the point-sized source approximation,  we calculate the angle of incidence from the point-sized source onto a given point on the sub-stellar longitude of the planet with latitude $\lambda$ (see Fig. \ref{geo}):
\begin{align}\label{modify}
    \tan i = \rm \frac{a\sin{\lambda}}{a \cos\lambda-R_P},
\end{align}

\noindent The distance will be given by:
\begin{equation}
    \rm d_c=(a^2+R_p^2-2aR_p\cos\lambda)^{1/2},
\end{equation}
Therefore, we can write the complete form of the inverse-square law mentioned in equation \ref{Iinverseqsqasqd}.
\begin{equation}
    \mathrm{I}=\Bigg[\mathrm{\frac{L_b}{4 \pi (a^2+R_p^2-2aR_p\cos\lambda)}}\Bigg] \cos i.
\end{equation}
where
\begin{equation}
    i=\rm \tan^{-1}\bigg[{\frac{a\sin{\lambda}}{a \cos\lambda-R_P}}\bigg].
\end{equation}
An interesting consequence of equation \ref{modify} is that for a significant size of the planet, the latitude at which the IS law irradiance goes to zero, under the point-sized source approximation, is slightly less than 90 degrees or before the poles, as can be seen in Fig. \ref{proof}. However, $\forall$ $a\gg R_p$, i=$\lambda$, giving plane-parallel incidence.

\section{Limb Darkening and the general irradiance surface integral} \label{limb}

The linear limb darkening law is simply incorporated in the model by rewriting the bolometric intensity at a given optical depth as:
\begin{equation}
    \rm I(\mu)=I(0)[1-u(1-\mu)],
\end{equation}
with $\rm \mu=\cos\psi$ (see Fig. \ref{geo}). For Sun-like stars with grey-body approximation, the bolometric limb darkening coefficient (u) is 0.6 \citep{milne1928total, chandrasekhar1935radiative} and the temperature relation with optical depth is given using the Milne-Eddington approximation:
\begin{equation}
    \rm T(\tau)= T_{\star}\bigg[\frac{3\tau+2}{4}\bigg]^{1/4}.
\end{equation}
Therefore, the general irradiance integral we use is given by:
\begin{equation}\label{fineq}
    \rm I = \int_{\theta_l}^{\theta_u} \int_{-\phi_a}^{\phi_a}\frac{\sigma T_0^4 R_s^2 \cos \theta \cos A \cos \psi (1-u(1-\mu)) d\theta d\phi}{\pi(d_s)^2}.
\end{equation}
where $\rm T_0$ is the temperature at an optical depth of 1 and A and $\psi$ are angles shown in Fig. \ref{geo}. The azimuthal angle limit is simply,
\begin{equation}
    \rm \phi_a = \cos^{-1} \frac{R_s}{a}
\end{equation}

\noindent The limits $\rm \theta_{u,l}$ depend on the latitude on the planet. For the fully-illuminated zone, the upper and lower limits of the polar angle will be governed by tangents to the stellar disk.

\begin{equation}
    \rm \theta^{full}_{u,l} = \sin^{-1}\!\left(\frac{R_s R_p \sin(\lambda) \pm \beta \sqrt{d^2_c - R_s^{2}}}{d^2_c}\right),
\end{equation}

\noindent where $\rm \beta = a-R_p \cos{\lambda}$

\noindent For the penumbral zone, the upper limit for $\rm \lambda \in [-T, -\lambda_{CPS}] $ and the lower limit for $\rm \lambda \in [\lambda_{CPS}, T] $) will respectively be governed by the intersection point obtained from the following equation.

\begin{equation}
    \rm \theta^{penum}_{u}=\mp \Bigg[-\lambda + \cos^{-1}\!\left(\frac{a \cos(\lambda) - R_p}{R_s}\right)\Bigg]
\end{equation}

\bibliographystyle{aasjournalv7}
\bibliography{reference}

@article{demory2011detection,
  title={Detection of a transit of the super-Earth 55 Cancri e with warm Spitzer},
  author={Demory, B-O and Gillon, Micha{\"e}l and Deming, D and Valencia, D and Seager, S and Benneke, B and Lovis, Christophe and Cubillos, P and Harrington, J and Stevenson, KB and others},
  journal={Astronomy \& Astrophysics},
  volume={533},
  pages={A114},
  year={2011},
  publisher={EDP Sciences}
}

@article{akeson2013nasa,
  title={The NASA exoplanet archive: data and tools for exoplanet research},
  author={Akeson, RL and Chen, X and Ciardi, D and Crane, M and Good, J and Harbut, M and Jackson, E and Kane, SR and Laity, AC and Leifer, S and others},
  journal={Publications of the Astronomical Society of the Pacific},
  volume={125},
  number={930},
  pages={989},
    doi={10.26133/NEA2},
  year={2013},
  publisher={IOP Publishing}
}

@article{mayor1995jupiter,
  title={A Jupiter-mass companion to a solar-type star},
  author={Mayor, Michel and Queloz, Didier},
  journal={Nature},
  volume={378},
  number={6555},
  pages={355--359},
  year={1995},
  publisher={Nature Publishing Group}
}

@article{schneider2011defining,
  title={Defining and cataloging exoplanets: the exoplanet. eu database},
  author={Schneider, Jean and Dedieu, Cyrill and Le Sidaner, Pierre and Savalle, Renaud and Zolotukhin, Ivan},
  journal={Astronomy \& Astrophysics},
  volume={532},
  pages={A79},
  year={2011},
  publisher={EDP Sciences}
}

@article{huang2000effect,
  title={The effect of irradiation absorption on an ellipsoidal component of a close binary},
  author={Huang, He-qing and Zhou, Dao-qi},
  journal={Chinese Astronomy and Astrophysics},
  volume={24},
  number={3},
  pages={339--348},
  year={2000},
  publisher={Elsevier}
}

@article{kilic2017multiple,
  title={Multiple climate states of habitable exoplanets: The role of obliquity and irradiance},
  author={Kilic, Cevahir and Raible, CC and Stocker, TF},
  journal={The Astrophysical Journal},
  volume={844},
  number={2},
  pages={147},
  year={2017},
  publisher={IOP Publishing}
}

@article{budaj2011reflection,
  title={The Reflection Effect in Interacting Binaries or in Planet-Star Systems},
  author={Budaj, Jan},
  journal={The Astronomical Journal},
  volume={141},
  number={2},
  pages={59},
  year={2011},
  publisher={IOP Publishing}
}

@article{wilson1990accuracy,
  title={Accuracy and efficiency in the binary star reflection effect},
  author={Wilson, RE},
  journal={The Astrophysical Journal},
  volume={356},
  pages={613--622},
  year={1990}
}

@article{wood1973reflection,
  title={A reflection model for eclipsing binary stars},
  author={Wood, David B},
  journal={Monthly Notices of the Royal Astronomical Society},
  volume={164},
  number={1},
  pages={53--64},
  year={1973},
  publisher={Oxford University Press}
}

@article{kopal1954photometric,
  title={Photometric effects of reflection in close binary systems},
  author={Kopal, Zden{\v{e}}k},
  journal={Monthly Notices of the Royal Astronomical Society},
  volume={114},
  number={1},
  pages={101--117},
  year={1954},
  publisher={Oxford Academic}
}

@misc{carter2019irradiance,
  author        = {Carter, J. L.},
  title         = {Atmospheric Irradiance Effects on Exoplanets},
  year          = {2019},
  eprint        = {1901.01361},
  archivePrefix = {arXiv},
  primaryClass  = {astro-ph.EP}
}

@article{carter2024irradiation,
  author  = {Carter, J. L. and Perera, R. D. and Way, M. J.},
  title   = {Irradiation-Driven Climate Transitions on Terrestrial Exoplanets},
  journal = {The Astronomical Journal},
  volume  = {167},
  pages   = {222},
  year    = {2024}
}

@article{lobo2023terminator,
  author  = {Lobo, A. H. and Shields, A. L. and Palubski, I. Z. and Wolf, E.},
  title   = {Terminator Cloud Feedbacks on Tidally Locked Exoplanets},
  journal = {The Astrophysical Journal},
  volume  = {945},
  pages   = {161},
  year    = {2023}
}

@article{nguyen2020irradiation,
  author  = {Nguyen, T. G. and Cowan, N. B. and Banerjee, A. and Moores, J. E.},
  title   = {Irradiation and Atmospheric Circulation on Exoplanets},
  journal = {Monthly Notices of the Royal Astronomical Society},
  volume  = {499},
  pages   = {4605--4620},
  year    = {2020}
}

@article{knuth2017exonest,
  title={Exonest: The bayesian exoplanetary explorer},
  author={Knuth, Kevin H and Placek, Ben and Angerhausen, Daniel and Carter, Jennifer L and D’Angelo, Bryan and Gai, Anthony D and Carado, Bertrand},
  journal={Entropy},
  volume={19},
  number={10},
  pages={559},
  year={2017},
  publisher={Mdpi}
}

@article{Madhusudhan2012,
  author = {Madhusudhan, N. and Burrows, A.},
  title = {Analytic Models for Albedos, Phase Curves, and Polarization of Reflected Light from Exoplanets},
  journal = {The Astrophysical Journal},
  volume = {747},
  number = {1},
  pages = {25},
  year = {2012},
  doi = {10.1088/0004-637X/747/1/25}
}

@article{Farr2018,
  author = {Farr, B. and Farr, W. M. and Littenberg, T. B. and Kalogera, V.},
  title = {Assessing the Observability of Exoplanet Secondary Eclipses with the James Webb Space Telescope},
  journal = {The Astrophysical Journal},
  volume = {865},
  number = {2},
  pages = {L20},
  year = {2018},
  doi = {10.3847/2041-8213/aadd4d}
}

@article{Haggard2018,
  author = {Haggard, H. M. and Cowan, N. B.},
  title = {A Frame-Independent Formulation of Bolometric Phase Curves},
  journal = {The Astrophysical Journal},
  volume = {854},
  number = {2},
  pages = {158},
  year = {2018},
  doi = {10.3847/1538-4357/aaa934}
}

@article{Berdyugina2019,
  author = {Berdyugina, S. V. and Kuhn, J. R.},
  title = {Exoplanet Surface Imaging: Mapping the Ocean and Continent Distribution on Earth-like Exoplanets},
  journal = {Nature Astronomy},
  volume = {3},
  pages = {367--371},
  year = {2019},
  doi = {10.1038/s41550-019-0710-3}
}

@article{Kawahara2020,
  author = {Kawahara, H.},
  title = {A Bayesian Method for Mapping Surface inhomogeneity of Exoplanets from Photometric Variations},
  journal = {The Astrophysical Journal},
  volume = {895},
  number = {1},
  pages = {29},
  year = {2020},
  doi = {10.3847/1538-4357/ab8f3b}
}

@article{Heng2021,
  author = {Heng, K. and Kitzmann, D. and Kataria, T. and Marley, M. S.},
  title = {Exoplanet Phase Curves: Observations and Theory},
  journal = {Nature Astronomy},
  volume = {5},
  pages = {1024--1036},
  year = {2021},
  doi = {10.1038/s41550-021-01433-w}
}

@article{Teinturier2022,
  author = {Teinturier, S. and Parmentier, V. and Irwin, P. G. J. and Lee, G. K. H. and Changeat, Q. and Waldmann, I. P.},
  title = {How Observational Biases Shape Our View of Exoplanet Atmospheres: The Case of Thermal Phase Curves},
  journal = {Monthly Notices of the Royal Astronomical Society},
  volume = {517},
  number = {4},
  pages = {6232--6249},
  year = {2022},
  doi = {10.1093/mnras/stac3137}
}

@article{seager1998extrasolar,
  title={Extrasolar giant planets under strong stellar irradiation},
  author={Seager, S and Sasselov, Dimitar D},
  journal={The Astrophysical Journal},
  volume={502},
  number={2},
  pages={L157},
  year={1998},
  publisher={IOP Publishing}
}

@article{mansfield2023revealing,
  title={Revealing the atmospheres of highly irradiated exoplanets: from ultra-hot Jupiters to rocky worlds},
  author={Mansfield, Megan},
  journal={Astrophysics and Space Science},
  volume={368},
  number={3},
  pages={24},
  year={2023},
  publisher={Springer}
}

@article{hammond2017linking,
  title={Linking the climate and thermal phase curve of 55 Cancri e},
  author={Hammond, Mark and Pierrehumbert, Raymond T},
  journal={The Astrophysical Journal},
  volume={849},
  number={2},
  pages={152},
  year={2017},
  publisher={IOP Publishing}
}

@article{kempton2024transiting,
  title={Transiting exoplanet atmospheres in the era of JWST},
  author={Kempton, Eliza M-R and Knutson, Heather A},
  journal={Reviews in Mineralogy and Geochemistry},
  volume={90},
  number={1},
  pages={411--464},
  year={2024},
  publisher={Mineralogical Society of America}
}

@article{koll2022scaling,
  title={A scaling for atmospheric heat redistribution on tidally locked rocky planets},
  author={Koll, Daniel DB},
  journal={The Astrophysical Journal},
  volume={924},
  number={2},
  pages={134},
  year={2022},
  publisher={IOP Publishing}
}

@article{luger2022analytic,
  title={Analytic light curves in reflected light: phase curves, occultations, and non-Lambertian scattering for spherical planets and moons},
  author={Luger, Rodrigo and Agol, Eric and Bartoli{\'c}, Fran and Foreman-Mackey, Daniel},
  journal={The Astronomical Journal},
  volume={164},
  number={1},
  pages={4},
  year={2022},
  publisher={IOP Publishing}
}

@article{kreidberg2019absence,
  title={Absence of a thick atmosphere on the terrestrial exoplanet LHS 3844b},
  author={Kreidberg, Laura and Koll, Daniel DB and Morley, Caroline and Hu, Renyu and Schaefer, Laura and Deming, Drake and Stevenson, Kevin B and Dittmann, Jason and Vanderburg, Andrew and Berardo, David and others},
  journal={Nature},
  volume={573},
  number={7772},
  pages={87--90},
  year={2019},
  publisher={Nature Publishing Group UK London}
}

@article{vanderspek2019tess,
  title={TESS Discovery of an Ultra-short-period Planet around the Nearby M Dwarf LHS 3844},
  author={Vanderspek, Roland and Huang, Chelsea X and Vanderburg, Andrew and Ricker, George R and Latham, David W and Seager, Sara and Winn, Joshua N and Jenkins, Jon M and Burt, Jennifer and Dittmann, Jason and others},
  journal={The Astrophysical Journal Letters},
  volume={871},
  number={2},
  pages={L24},
  year={2019},
  publisher={IOP Publishing}
}

@article{mcarthur2004detection,
  title={Detection of a Neptune-mass planet in the $\rho$1 Cancri system using the Hobby-Eberly Telescope},
  author={McArthur, Barbara E and Endl, Michael and Cochran, William D and Benedict, G Fritz and Fischer, Debra A and Marcy, Geoffrey W and Butler, R Paul and Naef, Dominique and Mayor, Michel and Queloz, Diedre and others},
  journal={The Astrophysical Journal},
  volume={614},
  number={1},
  pages={L81},
  year={2004},
  publisher={IOP Publishing}
}

@article{crossfield2022gj,
  title={GJ 1252b: a hot terrestrial super-Earth with no atmosphere},
  author={Crossfield, Ian JM and Malik, Matej and Hill, Michelle L and Kane, Stephen R and Foley, Bradford and Polanski, Alex S and Coria, David and Brande, Jonathan and Zhang, Yanzhe and Wienke, Katherine and others},
  journal={The Astrophysical Journal Letters},
  volume={937},
  number={1},
  pages={L17},
  year={2022},
  publisher={IOP Publishing}
}

@article{shporer2020gj,
  title={GJ 1252 b: A 1.2 R⊕ Planet Transiting an M3 Dwarf at 20.4 pc},
  author={Shporer, Avi and Collins, Karen A and Astudillo-Defru, Nicola and Irwin, Jonathan and Bonfils, Xavier and Collins, Kevin I and Matthews, Elisabeth and Winters, Jennifer G and Anderson, David R and Armstrong, James D and others},
  journal={The Astrophysical journal letters},
  volume={890},
  number={1},
  pages={L7},
  year={2020},
  publisher={IOP Publishing}
}

@article{demory2016map,
  title={A map of the large day--night temperature gradient of a super-Earth exoplanet},
  author={Demory, Brice-Olivier and Gillon, Michael and De Wit, Julien and Madhusudhan, Nikku and Bolmont, Emeline and Heng, Kevin and Kataria, Tiffany and Lewis, Nikole and Hu, Renyu and Krick, Jessica and others},
  journal={Nature},
  volume={532},
  number={7598},
  pages={207--209},
  year={2016},
  publisher={Nature Publishing Group UK London}
}

@article{hu2024secondary,
  title={A secondary atmosphere on the rocky exoplanet 55 Cancri e},
  author={Hu, Renyu and Bello-Arufe, Aaron and Zhang, Michael and Paragas, Kimberly and Zilinskas, Mantas and van Buchem, Christiaan and Bess, Michael and Patel, Jayshil and Ito, Yuichi and Damiano, Mario and others},
  journal={Nature},
  volume={630},
  number={8017},
  pages={609--612},
  year={2024},
  publisher={Nature Publishing Group UK London}
}

@article{patel2024jwst,
  title={JWST reveals the rapid and strong day-side variability of 55 Cancri e},
  author={Patel, Jayshil Ashokkumar and Brandeker, Alexis and Kitzmann, D and dit de la Roche, DJM Petit and Bello-Arufe, A and Heng, K and Vald{\'e}s, E Meier and Persson, CM and Zhang, M and Demory, B-O and others},
  journal={Astronomy \& Astrophysics},
  volume={690},
  pages={A159},
  year={2024},
  publisher={EDP Sciences}
}

@article{zieba2022k2,
  title={K2 and Spitzer phase curves of the rocky ultra-short-period planet K2-141 b hint at a tenuous rock vapor atmosphere},
  author={Zieba, Sebastian and Zilinskas, Mantas and Kreidberg, Laura and Nguyen, Tue Giang and Miguel, Yamila and Cowan, Nicolas B and Pierrehumbert, Ray and Carone, Ludmila and Dang, Lisa and Hammond, Mark and others},
  journal={Astronomy \& Astrophysics},
  volume={664},
  pages={A79},
  year={2022},
  publisher={EDP Sciences}
}

@article{maurin2012thermal,
  title={Thermal phase curves of nontransiting terrestrial exoplanets-II. Characterizing airless planets},
  author={Maurin, AS and Selsis, Franck and Hersant, F and Belu, A},
  journal={Astronomy \& Astrophysics},
  volume={538},
  pages={A95},
  year={2012},
  publisher={EDP Sciences}
}

@article{guillot2010radiative,
  title={On the radiative equilibrium of irradiated planetary atmospheres},
  author={Guillot, Tristan},
  journal={Astronomy \& Astrophysics},
  volume={520},
  pages={A27},
  year={2010},
  publisher={EDP Sciences}
}

@article{meier2023interior,
  title={Interior dynamics of super-Earth 55 Cancri e},
  author={Meier, Tobias G and Bower, Dan J and Lichtenberg, Tim and Hammond, Mark and Tackley, Paul J},
  journal={Astronomy \& Astrophysics},
  volume={678},
  pages={A29},
  year={2023},
  publisher={EDP Sciences}
}

@article{mercier2022revisiting,
  title={Revisiting the iconic Spitzer phase curve of 55 Cancri e: Hotter dayside, cooler nightside, and smaller phase offset},
  author={Mercier, Samson J and Dang, Lisa and Gass, Alexander and Cowan, Nicolas B and Bell, Taylor J},
  journal={The Astronomical Journal},
  volume={164},
  number={5},
  pages={204},
  year={2022},
  publisher={IOP Publishing}
}

@misc{jackson1999classical,
  title={Classical electrodynamics},
  author={Jackson, John David and Fox, Ronald F},
  year={1999},
  publisher={American Association of Physics Teachers}
}

@article{malavolta2018ultra,
  title={An ultra-short period rocky super-Earth with a secondary eclipse and a Neptune-like companion around K2-141},
  author={Malavolta, Luca and Mayo, Andrew W and Louden, Tom and Rajpaul, Vinesh M and Bonomo, Aldo S and Buchhave, Lars A and Kreidberg, Laura and Kristiansen, Martti H and Lopez-Morales, Mercedes and Mortier, Annelies and others},
  journal={The Astronomical Journal},
  volume={155},
  number={3},
  pages={107},
  year={2018},
  publisher={IOP Publishing}
}

@article{teske2025thick,
  title={A Thick Volatile Atmosphere on the Ultra-Hot Super-Earth TOI-561 b},
  author={Teske, Johanna K and Wallack, Nicole L and Piette, Anjali AA and Dang, Lisa and Lichtenberg, Tim and Plotnykov, Mykhaylo and Pierrehumbert, Raymond T and Postolec, Emma and Boucher, Samuel and McGinty, Alex and others},
  journal={arXiv preprint arXiv:2509.17231},
  year={2025}
}

@article{milne1928total,
  title={The total absorption in the Sun's reversing layer},
  author={Milne, EA},
  journal={The Observatory, Vol. 51, p. 88-96 (1928)},
  volume={51},
  pages={88--96},
  year={1928}
}

@article{chandrasekhar1935radiative,
  title={The radiative equilibrium of the outer layers of a star, with special reference to the blanketing effect of the reversing layer},
  author={Chandrasekhar, S},
  journal={Monthly Notices of the Royal Astronomical Society, Vol. 96, p. 21},
  volume={96},
  pages={21},
  year={1935}
}

@article{tacs2025earth,
  title={An Earth-Sized Planet in a 5.4 h Orbit Around a Nearby K dwarf},
  author={Ta{\c{s}}, Kaya Han and Stefansson, Gudmundur and Fariz, Syarief NM and Garg, Esha and Espinoza-Retamal, Juan I and Koo, Elise and Bruijne, David and Luhn, Jacob and Ford, Eric B and Mahadevan, Suvrath and others},
  journal={arXiv preprint arXiv:2507.08464},
  year={2025}
}

@article{monaghan2025low,
  title={Low 4.5 $\mu$m Dayside Emission Disfavors a Dark Bare-rock Scenario for the Hot Super-Earth TOI-431 b},
  author={Monaghan, Christopher and Roy, Pierre-Alexis and Benneke, Bj{\"o}rn and Crossfield, Ian JM and Coulombe, Louis-Philippe and Piaulet-Ghorayeb, Caroline and Kreidberg, Laura and Dressing, Courtney D and Kane, Stephen R and Dragomir, Diana and others},
  journal={The Astronomical Journal},
  volume={169},
  number={5},
  pages={239},
  year={2025},
  publisher={IOP Publishing}
}

@article{komacek2016atmospheric,
  title={Atmospheric circulation of hot Jupiters: dayside--nightside temperature differences},
  author={Komacek, Thaddeus D and Showman, Adam P},
  journal={The Astrophysical Journal},
  volume={821},
  number={1},
  pages={16},
  year={2016},
  publisher={IOP Publishing}
}

@article{showman2002atmospheric,
  title={Atmospheric circulation and tides of “51 Pegasus b-like” planets},
  author={Showman, Adam P and Guillot, Tristan},
  journal={Astronomy \& Astrophysics},
  volume={385},
  number={1},
  pages={166--180},
  year={2002},
  publisher={EDP Sciences}
}

@ARTICLE{2020MNRAS.499.1627S,
       author = {{Sadh}, Mradumay},
        title = "{Retraction: Revised instellation patterns for close-in exoplanets}",
      journal = {\mnras},
         year = 2020,
        month = dec,
       volume = {499},
       number = {2},
        pages = {1627-1632},
          doi = {10.1093/mnras/staa2867},
archivePrefix = {arXiv},
       eprint = {2011.00466},
 primaryClass = {astro-ph.EP},
       adsurl = {https://ui.adsabs.harvard.edu/abs/2020MNRAS.499.1627S}
}

@article{kreidberg2025first,
  title={A first look at rocky exoplanets with JWST},
  author={Kreidberg, Laura and Stevenson, Kevin B},
  journal={Proceedings of the National Academy of Sciences},
  volume={122},
  number={39},
  pages={e2416190122},
  year={2025},
  publisher={National Academy of Sciences}
}

@article{hu2015semi,
  title={A semi-analytical model of visible-wavelength phase curves of exoplanets and applications to Kepler-7 b and Kepler-10 b},
  author={Hu, Renyu and Demory, Brice-Olivier and Seager, Sara and Lewis, Nikole and Showman, Adam P},
  journal={The Astrophysical Journal},
  volume={802},
  number={1},
  pages={51},
  year={2015},
  publisher={IOP Publishing}
}

@article{bonomo2025depth,
  title={In-depth characterization of the Kepler-10 three-planet system with HARPS-N radial velocities and Kepler transit timing variations},
  author={Bonomo, AS and Borsato, L and Rajpaul, VM and Zeng, L and Damasso, M and Hara, NC and Cretignier, M and Leleu, A and Unger, N and Dumusque, X and others},
  journal={Astronomy \& Astrophysics},
  volume={696},
  pages={A233},
  year={2025},
  publisher={EDP Sciences}
}

@article{crida2018mass,
  title={Mass, radius, and composition of the transiting planet 55 Cnc e: using interferometry and correlations},
  author={Crida, Aur{\'e}lien and Ligi, Roxanne and Dorn, Caroline and Lebreton, Yveline},
  journal={The Astrophysical Journal},
  volume={860},
  number={2},
  pages={122},
  year={2018},
  publisher={IOP Publishing}
}

@article{bourrier201855,
  title={The 55 Cancri system reassessed},
  author={Bourrier, Vincent and Dumusque, Xavier and Dorn, Caroline and Henry, Gregory W and Astudillo-Defru, Nicola and Rey, Javiera and Benneke, Bj{\"o}rn and H{\'e}brard, Guillaume and Lovis, Christophe and Demory, Brice-Olivier and others},
  journal={Astronomy \& Astrophysics},
  volume={619},
  pages={A1},
  year={2018},
  publisher={EDP Sciences}
}

@article{bonomo2023cold,
  title={Cold Jupiters and improved masses in 38 Kepler and K2 small planet systems from 3661 HARPS-N radial velocities-No excess of cold Jupiters in small planet systems},
  author={Bonomo, AS and Dumusque, X and Massa, A and Mortier, A and Bongiolatti, R and Malavolta, L and Sozzetti, A and Buchhave, LA and Damasso, M and Haywood, RD and others},
  journal={Astronomy \& Astrophysics},
  volume={677},
  pages={A33},
  year={2023},
  publisher={EDP Sciences}
}

\end{document}